# CCD PHOTOMETRY OF VARIABLE STARS IN THE GLOBULAR CLUSTER RU 106 [1]

Janusz Kaluzny

Warsaw University Observatory, Al. Ujazdowskie 4, 00-478 Warsaw, Poland
e-mail: jka@sirius.astrouw.edu.pl

and

Wojciech Krzeminski

Carnegie Observatories, Las Campanas Observatory, Casilla 601, LaSerena, Chile
e-mail: wojtek@roses.ctio.noao.edu

and

Beata Mazur

Copernicus Astronomical Center, ul. Bartycka 18, Warsaw, Poland
e-mail: batka@camk.edu.pl

## ABSTRACT

BV photometry is presented for 12 RR Lyr variables discovered in the presumably young galactic globular cluster Ruprecht 106. All variables are type RRab, and their periods span a narrow range from 0.574 to 0.652 day. We report also on the discovery of 3 SX Phe variables among the cluster blue stragglers. A likely background RR Lyr variable and two foreground contact binaries were also found in the cluster field. The reddening of Ruprecht 106 is estimated at $E(B-V) = 0.20$ based on the (B-V) colors exhibited by the cluster RR Lyr variables at minimum light.

Analysis of the period" versus "amplitude" and "period" versus "rise-time" diagrams suggests similar metallicities of Ruprecht 106 and M3. A peak (or bump) is present in the luminosity function of the red giant branch of Ru 106. Its position relative to the horizontal branch is consistent with a cluster metallicity of [Fe/H] $\geq -1.6$.

*Subject headings:* clusters: globular — stars: variables – color magnitude diagram

---

## 1. Introduction

Ruprecht 106 (hereafter Ru 106) belongs to a small group of relatively young galactic globular clusters. Buonanno *et al.* (1990) and Buonanno *et al.* (1993) published color-magnitude diagrams of Ru 106 which were used to constrain cluster parameters. They found that the metallicity of Ru 106 is [Fe/H] = −1.9 ± 0.2 and that it is 3–5 Gyr younger than other galactic globular clusters with similar metallicities. It has been suggested that Ru 106, as well as some other exceptionally young globulars, may have been captured by the Milky Way from the Magellanic Clouds (Lin & Richer 1992). However, such a hypothesis has been criticized by van den Bergh (1994).

In 1991 two of us (WK & BM) used the 2.5-m duPont telescope at Las Campanas Observatory to conduct a preliminary survey for variable blue stragglers in Ru 106. This survey resulted in serendipitious discovery of several candidates for RR Lyr variables. In this paper we present results of follow up observations aimed at a study of the photometric properties of RR Lyr stars in Ru 106. An independent observing run conducted on the duPont telescope was devoted to a detailed study of variable blue stragglers in this cluster. Results of these observations will be published in a complementary paper (Krzeminski *et al.*, in preparation).

## 2. Observations and data reductions

A field centered approximately on the cluster center was monitored during 5 nights spanning the period from April 27 to May 01, 1993 (UT). All observations were made using the 1-m Swope telescope at Las Campanas Observatory. A thinned $1024 \times 1024$ Tektronix chip with a scale of 0.695 arcsec/pixel, was used as the detector. Preliminary processing of the CCD frames was done with the standard routines in the IRAF-CCDPROC[2] package. The flat-field frames were prepared by combining "dome flats" and exposures of the twilight sky. The reduction procedures reduced total instrumental systematics to below 1% for the central $780 \times 780$ pixels$^2$ area of the images. Some systematic residual pattern at the 1%-4% level was left near borders of the images, perhaps due to an uneven antireflecting coating.

Observations were performed using Johnson B and V filters. The exposure time ranged from 320 to 400 sec for the V-band, and was equal to 500 sec for the B-band. The observations in both filters were performed on 3 nights out of 5. Each observation in the B filter was followed by two exposures in the V filter. On two nights observations were performed only in the V filter. A full listing of all exposures is given in Table 1 which is available on the CD-ROM supplemented to the *Astronomical Journal*.

---

[2]IRAF is distributed by the National Optical Astronomical Observatories, operated by the Association of Universities for Research in Astronomy, Inc., under contract with the National Science Foundation.

Several fields containing standard stars from Landolt (1992) were observed during 3 nights. Independent transformations from the instrumental to the standard BV system were first derived for these nights. Subsequently we adopted averaged values for the color terms and derived new zero points and extinction coefficients for each night. Photometry of Ru 106 was transformed to the BV system using data obtained on the night of May 1, 1993. The following relations were obtained for that night:

$$v = const + V - 0.019(B - V) - 0.168X \tag{1}$$

$$(b - v) = const + 0.913(B - V) - 0.085X \tag{2}$$

where $X$ is the airmass, and lower-case letters refer to the instrumental magnitudes. In Fig. 1 we show residuals between the standard and the recovered magnitudes and colors for 12 stars from 3 Landolt (1992) fields which were observed immediately before Ru 106 on the night of May 1, 1993. These 3 fields were observed at air masses ranging from 1.17 to 1.86.

A total of 232 V and 54 B images were used for photometry of Ru 106. The instrumental photometry was extracted using DAOPHOT/ALLSTAR package (Stetson 1987, 1991). It was found that the point spread function (PSF) showed some positional dependence, and a PSF varying linearly across the image was adopted.

The observed field was rather crowded. Moreover, the analyzed images were poorly sampled due to the large scale of pixels. The FWHM reached 2.0 pixels on the best images. To improve the stability of the derived photometry, we adopted the following procedure. First, we extracted photometry from three V frames, which were judged to be of a particularly high quality. Coordinates of all measured stars were transformed to one common system and a master list containing objects detected on at least two frames was created. A list of stars suitable for calculation of the PSF was also created. Reduction of individual frames started with a derivation of the aperture photometry for the relatively bright stars. These stars were then used to transform the coordinates of stars from the master list to the coordinates of the currently processed frame. Subsequently, aperture photometry was derived for stars from the master list and stars used for determination of the PSF were identified. After calculating the PSF for the current frame we ran the ALLSTAR program (Stetson 1987) to obtain profile photometry for stars from the master list. Such an approach not only saves CPU time but, more importantly, helps to obtain better photometry from relatively poor frames affected by bad seeing or bright sky. A similar procedure was used for the B-band images. Photometry for all frames was transformed to a common instrumental system and finally transformed to the BV system. Data bases for the V and B observations were created and the average magnitudes for stars from the data bases were used to create the color-magnitude diagram (CMD hereafter) for the monitored field. 5206 and 5208 stars were included in the V and B data bases, respectively.

## 3. Variable Stars

A search for variables was conducted using the data bases for both filters. For every star which was measured on at least 50% of frames for a given data set, we calculated a $\chi^2$ statistic. Objects with $P(\chi^2) < 10^{-4}$ (eg. Press *et al.* 1986) were considered candidate variables and their light curves were tested for variability with periods in the range 0.03 to 5 days. To determine the most probable periods we used an *aov* statistic (Schwarzenberg-Czerny 1989, 1991). To look more efficiently for faint short-period variables we analyzed separately the data from the first two nights of the run. Photometry obtained on these two nights was particularly deep and accurate due to favorable observing conditions (dark sky and good seeing). In Fig. 2 we present a plot of *rms* deviation versus the average V magnitude for the light curves which are based on frames taken during the first two nights of the observing run. The limiting magnitude of the photometry is $V \approx 22.3$. The median value of *rms* is about 0.02 at $V = 19$ and reaches about 0.05 at $V = 20.7$.

We discovered a total of 18 periodic variables located in the field of Ru 106. Their equatorial and rectangular coordinates are listed in Table 2. The rectangular coordinates correspond to the V-band image which is available on the CD-ROM supplemented to the *Astronomical Journal*. A transformation from rectangular to equatorial coordinates was derived based on positions of 16 stars from the Guide Star Catalogue (Lasker *et al.* 1988).

Our sample of identified variables consists of 13 RR Lyr stars, 2 contact binaries and 3 SX Phe stars. Phased light curves of all variables are shown in Figs. 3 and 4. We did not make any tests to determine the degree of completeness of our photometry of Ru 106. Based on the overall quality of the photometry we are inclined to believe that our sample of RR Lyr variables is complete down to amplitudes of about 0.1 mag. On the other hand, it is likely that we missed some variable blue stragglers.

In the following discussion we focus our attention mostly on the RR Lyr variables. Particularly, we use the observed characteristics of these stars to constrain the reddening and metallicity of Ru 106. A dedicated observing session was devoted to study variable blue stragglers hosted by the cluster. Results of that study will be published elsewhere (Krzeminski *et al.*, in preparation).

### 3.1. RR Lyr variables

Table 3 lists the basic characteristics of the light curves of 13 RR Lyr stars discovered in the field of Ru 106. The mean B and V magnitudes were calculated by numerically integrating the phased light curves after converting them into an intensity scale. Mean values for $(B - V)$ were obtained by integrating the color curves. Table 3 also lists the B and V amplitudes of the light curves. V-band amplitudes could be derived for all variables. B-band amplitudes could not be derived for stars V10 and V15 due to incomplete coverage of their light curves. The color at the minimum light is denoted by $(B - V)_{min}$. It was derived by averaging observations during the interval $0.50 < \phi < 0.80$ (we follow Sturch's (1966) definition). The parameter $\Delta\phi$ gives the fraction of the period from minimum to maximum light. It was calculated using the V-band light curves.

In Fig. 5 we show a section of the color-magnitude diagram for the central part of Ru 106. Variable V2 is located about 1.3 mag below the horizontal branch of the cluster. This suggests that it is a field object located behind the cluster. Variable V2 was not included in the derivation of cluster reddening and metallicity.

The properties of the Ru 106 RR Lyr stars can be used to estimate some basic parameters of the cluster itself. In Fig. 6 we show the positions of the variables on a "period" versus "B amplitude" diagram. Fig. 7 shows "period" versus "rise-time" $\Delta\phi$. Fiducial relations for RRab stars from globular clusters M3 and M15 are also marked in Figs. 6 and 7. These relations were adopted after Sandage *et al.* (1981). The "reference clusters" M3 and M15 have, respectively, $[Fe/H] = -1.66$ and $-2.15$ (Zinn & West 1984). The Ru 106 variables are grouped in Figs. 6 and 7 around relations defined by the M3 stars. In fact, the majority of Ru 106 stars fall on the metal-rich side of the M3 line indicating that the metallicity of Ru 106 is likely to be $[Fe/H] \geq -1.66$. This is consistent with the results of Da Costa *et al.* (1992) who obtained $[Fe/H] = -1.69 \pm 0.05$ from analysis of the spectra of 7 Ru 106 giants. Sarajedini (1994) derived $[Fe/H] = -1.61 \pm 0.20$ from an analysis of the V/V-I CMD of the cluster.

Sturch (1966) presented a formula relating the unreddened colors of RRab variables with their periods and metallicities. To calculate the $E(B - V)$ for the Ru 106 variables, we used Sturch's formula in the form given by Walker (1990):

$$E(B - V) = (B - V)_{min} - 0.24P - 0.056[Fe/H] - 0.347. \qquad (3)$$

This equation holds when the Zinn & West (1984) metallicity scale is used. Individual values of the reddening were calculated for 10 variables after adopting [Fe/H] $= -1.66$. Stars V5 and V9 were dropped from the analysis. For V9 the value of $(B - V)_{min}$ is uncertain while $E(B - V)$ derived for V5 deviates significantly from the average value obtained for the remaining variables. The derived individual values of $E(B - V)$ range from 0.167 to 0.223 and the average value for the 10 variables is $< E(B - V) >= 0.183 \pm 0.002$. Knowing the reddening we are in position to estimate the cluster metallicity from the color of its red giant branch. Zinn and West (1984) calibrate the intrinsic color of the giant branch at the level of the horizontal branch as:

$$(B - V)_{0,g} = 1.16 + 0.23[\text{Fe/H}]. \tag{4}$$

Using the data shown in Fig. 5 we estimate $(B - V)_g = 0.93 \pm 0.025$. The quoted error includes both uncertainty of the zero-point of the color transformation, and the uncertainty involved in the determination of the fiducial relation for the RGB of Ru 106. It is encouraging that the same value of $(B - V)_g$ comes from the fiducial sequence provided by Buonanno $et$ $al.$ (1993). For $E(B - V) = 0.18$ we obtain $(B - V)_{0,g} = 0.75$ and a cluster metallicity [Fe/H] $= -1.78 \pm 0.11$ follows from Eq. 4. It has to be kept in mind that clusters used by Zinn & West to calibrate Eq. 4 show a scatter of $\sigma = 0.20$ around the mean relation. The derived metallicity is by 0.12 lower than the value which was used above to calculate $E(B - V)$ from Eq. 3. In fact Eqs. 3 and 4 can be solved simultaneously yielding a self-consistent values of $E(B - V)$ and [Fe/H]. Such a procedure gives $E(B - V) = 0.198 \pm 0.002$ and [Fe/H] $= -1.87 \pm 0.23$.

Based on the properties of their RR Lyr stars, galactic globular clusters are conventionally classified into two groups: Oosterhoff type I (metal-intermediate objects with $< P_{ab} >\approx 0.55$ and a predominance of ab over c-type RR Lyr variables) and Oosterhoff type II (metal-poor objects with $< P_{ab} >\approx 0.65$ and comparable fractions of c-type and ab-type variables). Recent reviews of this subject and a summary of relevant observational data can be found in Sandage (1993) and in Bono $et$ $al.$ (1994). The average period of the 12 RRab variables discovered in Ru 106 is $< P_{ab} >= 0.616$ with $\sigma_{ab} = 0.021$. We note that the distribution of periods of the Ru 106 variables is extremely narrow. None of clusters listed in Table 1 of Sandage (1993) has such a small value of $\sigma_{ab}$ as Ru 106. Our data also show a complete lack of RRc variables in Ru 106. Only 4 out of 37 clusters listed by Sandage (1993) are devoid of RRc stars. Sandage (1993) devised a method allowing a correctionof the observed mean period values $< P_{ab} >$ to what they would have been if the distribution of variables for a given cluster had been uniform. He obtained:

$$\Delta log < P >= -0.0267(r - 1)/(r + 1), \tag{5}$$

where $r$ is the ratio of the number of HB stars at the red edge to stars at the blue edge. For Ru106 we may adopt $\Delta log < P >= -0.027$ as its HB lacks any stars on the blue side of the instability strip. The corrected mean period is then $log < P_{ab} >_{corr} = -0.237$. In Fig. 8 we show relation between corrected average period $< P_{ab} >$ and metallicity which was adopted by Sandage (1993). Positions of clusters used to calibrate that relation are shown together with positions of Ru 106 plotted for [Fe/H] $= -1.9 \pm 0.2$ (Buonanno *et al.* 1993) and for [Fe/H] $= -1.69 \pm 0.05$ (Da Costa *et al.* 1992). The relatively low average period of RRab stars in Ru 106 is consistent with the higher metallicity of the cluster. A formal application of Sandage's calibration:

$$log < P_{ab} >_{corr} = -0.121 \times [Fe/H] - 0.431 \qquad (6)$$

leads to [Fe/H] $= -1.60$. When deriving his calibration (Eq. 5 above), Sandage (1993) dropped from consideration clusters with $-1.9 < [Fe/H] < -1.7$. Clusters from this metallicity range tend to have extremely blue horizontal branches. Their RR Lyr variables are on the evolved track on their way to the AGB. Such variables are brighter than the ZAHB and have longer periods than variables on the ZAHB (see also Lee *et al.* 1990). Clearly this problem is of no concern in the case of Ru 106 as it possesses a red HB.

## 3.2. Contact binaries and SX Phe stars

Buonanno *et al.* (1990) noted that Ru 106 hosts a sizeable population of blue stragglers. Two types of variables are known to occur among blue stragglers in globular clusters. These are eclipsing binaries (Hut *et al.* 1992)) and pulsating stars of SX Phe type (Nemec *et al.* 1994). Table 4 gives some basic photometric data for the SX Phe stars and contact binaries discovered in the field of Ru 106. The positions of these variables in the cluster CMD are shown in Fig. 9. Both contact binaries are too bright to be cluster members, and are likely foreground stars. All three SX Phe stars are located in the area of the CMD occupied by cluster blue stragglers and these variables are very likely cluster members. It has to be noted that our photometry of blue stragglers in Ru 106 is rather poor. These stars are relatively faint and most of them are located in the crowded central part of the cluster. The light curves of SX Phe stars obtained in the course of this survey are slightly distorted due to relatively long exposure times as compared with periods of pulsation of the variables. More detailed study of variable blue stragglers in Ru 106 will be published elsewhere (Krzeminski *et al.* in preparation).

## 4. The color-magnitude diagram

Photometry extracted from 232 images in the V band and from 54 images in the B band was averaged to produce a color magnitude diagram (CMD) for the observed field. Only stars measured on at least 130 V images and on at least 30 B images were retained in the final CMD. No attempt was made to estimate degree of incompleteness of the photometry. The limiting magnitude of our BV photometry is $V_{lim} \approx 22.2$ for stars located on the cluster main sequence. Deeper photometry of Ru 106 was published by Buonanno *et al.* (1993; $V_{lim} \approx 23.7$) and Buonanno *et al.* (1992; $V_{lim} \approx 22.5$). Data presented here cover, however, a much larger field than in the quoted studies.

In Fig. 10 we show the CMD for an $8.6 \times 8.6$ arcmin$^2$ field centered approximately on the cluster center. The field monitored for variables had a size of $11.6 \times 11.6$ arcmin$^2$. However, as we reported in Sec. 2, no precise photometry could be obtained for stars located near the edges of the images. A large fraction of stars in Fig. 10 are field objects. Despite this complication two interesting features of the CMD can easily be noted. First, the horizontal branch of the cluster lacks any stars located to the blue of the RR Lyr variables. Second, the blue edge of the cluster main sequence is sharply marked below the turnoff region. This indicates that the metallicity of the background halo population is not lower than the metallicity of Ru 106.

Two very blue objects with $B - V < -0.1$ are present in Fig. 10. These stars are located at $V \approx 20.3$ and $V \approx 22.0$. If they are cluster members or background objects then their unreddened colors are $B - V < -0.3$ for the adopted $E(B - V) = 0.20$. The fainter of these blue stars is located at an angular distance $r \approx 1.9$ arcmin from the cluster center, while the corresponding distance for the brighter star is $r \approx 3.4$ arcmin. These distances are lower than the tidal radius of the cluster. However, the available data do not allow to reject possibility that both identified blue objects are field stars or unresolved extra-galactic objects.

The most remote of the RR Lyr variables belonging to Ru 106 is the star V1, with a projected distance from the cluster center of 4.8 arcmin. This sets a lower limit on the tidal radius of Ru 106.

All 3 SX Phe stars reported in this paper are located closer that 2 arcmin from the cluster center. The group of potential blue stragglers visible in the CMD of Ru 106 extends down to $V \approx 20.9$. To look in more details on the blue stragglers population of the cluster we isolated two groups of stars: those lying within radius $R = 2.94$ arcmin from the cluster center, and those lying in an outer ring, between 3.31 and 4.43 arcmin. The inner circle and the outer ring cover equal areas on the sky. The CMDs for both selected subfields are presented in Fig. 11. We limited our attention to stars with $V < 21.7$ to minimize problems related to the incompleteness of the photometry whose degree depends heavily on the distance from the cluster center. As we noted above, the radius of Ru 106 exceeds 4.8 arcmin. Therefore, the CMD for the outer ring is populated by some cluster members. In fact a clearly marked main sequence of Ru 106 is visible on the right panel of Fig. 11. However, at the same time only two blue straggler candidates are visible in the CMD for the outer ring. We used data presented in Fig. 11 to obtain a field-star corrected CMD for the inner region of Ru 106. The outer ring served as a "comparison field". For each star from the "comparison field" a nearest match in the CMD for the inner field was located. Subsequently, a pair of stars with the lowest separation was removed from both corresponding CMDs. This procedure was continued as long as it was possible to locate pairs with a separation in magnitude $\delta V < 0.35$ and a separation in color $\delta (B-V) < 0.20$. The resulting "cleaned" CMD for the inner region of Ru 106 is shown in Fig. 12. Obviously the applied cleaning procedure was not very rigorous. However, some interesting conclusions can be made based on the data shown in Fig. 12. First, it may be noted that the population of blue stragglers extends toward the red colors well beyond the cluster turnoff. Such stars are sometimes called yellow stragglers. The group of yellow stragglers terminates about 0.75 mag above the top of the cluster main sequence. This is consistent with the hypothesis that yellow stragglers are binaries composed of turnoff stars. Our data show 38 blue stragglers with $(B-V) < 0.50$ and $19 < V < 20.3$. This number includes also 3 variable blue stragglers which are not marked in Fig. 12. A more detailed study of the blue/yellow straggler population in Ru 106 will be presented by Krzeminski *et al.* (in preparation).

The second interesting feature visible in Fig. 12 is the apparent clump occurring on the red giant branch, and centered at $V \approx 17.5$. Stellar models predict that the differential luminosity function of the red giants in globular clusters should show a small "bump". In fact such bumps have been observed in several globulars (Fusi Pecci *et al.* 1990, King *et al.* 1985). The position in luminosity of the RGB bump is a function of metal abundance, helium abundance and cluster age. Fusi Pecci *et al.* (1990) defined parameter $\Delta V_{bump}^{HB}$, measuring the difference in magnitudes between the horizontal branch and the faint edge of the clump. The relation between [Fe/H] and $\Delta V_{bump}^{HB}$ for 11 cluster analyzed by Fusi Peci *et al.* (1990) is shown in Fig. 13. For Ru 106 we estimate $\Delta V_{bump}^{HB} = -0.17$. This value indicates a relatively high metallicity of the cluster and is consistent with our assumed metallicity of Ru 106 of [Fe/H] $\approx -1.6$ (see Sec. 3.1). $\Delta V_{bump}^{HB}$ is also a function of cluster age. For fixed metallicity, $\Delta V_{bump}^{HB}$ increases with increasing age. Hence, we are likely to underestimate the metallicity of Ru 106 by comparing it in Fig. 13 with supposedly older clusters.

## 5. Summary

We summarize the essential results of this paper:

1. We have obtained 232 V and 54 B CCD images of the field of Ru 106 and have discovered 18 variable stars. Two of these stars are foreground contact binaries and one is an RR Lyr variable probably located behind the cluster. Three of the variables are SX Phe stars belonging to the population of Ru 106 blue stragglers. The remaining 12 variables are RR Lyr stars that appear to be members of the cluster. All RR Lyr stars identified in the field of Ru 106 belong to RRab subtype. Their periods span a narrow range from 0.574 to 0.652 days with the average value $< P_{ab} >= 0.616$.

2. The location of the Ru 106 RR Lyr stars on the "period" versus "amplitude" and "period" versus "rise-time" diagram indicates that the cluster metallicity is similar to that of M3.

3. The minimum-light colors of the RRab stars in Ru 106 imply a cluster reddening of $E(B - V) = 0.183 \pm 0.002$ for the adopted $[Fe/H] = -1.66$. The unreddened color of the red giant branch $(B - V)_{0,g} = 0.75$ leads then to an estimated metallicity of the cluster of $[Fe/H] = -1.78 \pm 0.23$. By requiring a self-consistent values of the reddening and metallicity we obtained $E(B - V) = 0.198 \pm 0.002$ and $[Fe/H] = -1.87 \pm 0.11$. Finally, using the Sandage's (1993) relation between $< P_{ab} >_{corr}$ and metallicity we obtained $[Fe/H] = -1.60$. It may be concluded that the observed characteristics of RRab stars in Ru 106 favor a relatively high metallicity of the cluster.

4. A CMD for 3535 stars from a field of size $8.6 \times 8.6$ arcmin$^2$ centered on the cluster center was presented. We confirm the earlier reports about a rich population of blue stragglers hosted by Ru 106. We noted however, that the cluster hosts also a sizeable population of yellow stragglers.

5. We identified an apparent "clump" on the red giant branch of the cluster. Position of this "clump", relatively to the horizontal branch, implies $[Fe/H] \geq -1.6$ for Ru 106.

6. It was found that the tidal radius of the cluster exceeds 4.8 arcmin.

We thank Ian Thompson and Kazik Stepien for helpful comments on the early version of this paper. JK was supported by the KBN grant 2P03D-008-08 and BM was supported by the KBN grant 2P304-013-07. JK and BM wishes to acknowledge with gratitude the kind hospitality of the Las Campanas Observatory during their stays in Chile.

# Appendix

Tables 5–13 list the individual observations for all of the newly discovered variables in Ru 106. These tables are presented in their complete form on the CD-ROM supplemented to the *Astronomical Journal*. The top of the first page of Table 5 is printed to illustrate the format.

Table 14 lists the BV photometry for 3535 stars from the field of Ru 106. This table is presented in its complete form on the CD-ROM supplemented to the *Astronomical Journal*. The top of the first page of Table 14 is printed to illustrate the format.

Table 2: Rectangular and equatorial coordinates for variables identified in the field of Ru 106. The X and Y coordinates correspond to positions on the image available on the CD-ROM supplemented to the *Astronomical Journal*. The seventh column gives the time of maximum light for the RR Lyr and SX Phe variables and the time of minimum light for the EW variables.

| ID | X | Y | RA(2000) | Dec(2000) | P | $HJD_0$ | Type |
|-----|-----|-----|------------|-----------|---------|-----------|------|
|     |     |     |            |           | days    | 2449100+  |      |
| V1  | 142 | 293 | 12:38:24.3 | -51:04:59 | 0.6104  | 4.6073    | RRab |
| V2  | 183 | 740 | 12:38:57.3 | -51:05:24 | 0.6268  | 4.9788    | RRab |
| V3  | 393 | 588 | 12:38:46.3 | -51:07:52 | 0.03952 | 4.5534    | SX   |
| V4  | 419 | 375 | 12:38:30.5 | -51:08:11 | 0.04900 | 4.5767    | SX   |
| V5  | 455 | 587 | 12:38:46.2 | -51:08:34 | 0.6006  | 5.0388    | RRab |
| V6  | 455 | 379 | 12:38:30.8 | -51:08:36 | 0.6408  | 5.1872    | RRab |
| V7  | 490 | 350 | 12:38:28.7 | -51:09:00 | 0.27552 | 4.8918    | EW   |
| V8  | 522 | 559 | 12:38:44.2 | -51:09:21 | 0.6208  | 4.5773    | RRab |
| V9  | 527 | 505 | 12:38:40.2 | -51:09:25 | 0.5740  | 5.0801    | RRab |
| V10 | 537 | 490 | 12:38:39.1 | -51:09:32 | 0.6019  | 4.8773    | RRab |
| V11 | 538 | 616 | 12:38:48.4 | -51:09:32 | 0.04777 | 4.5688    | SX   |
| V12 | 548 | 510 | 12:38:40.6 | -51:09:39 | 0.6221  | 4.8398    | RRab |
| V13 | 563 | 675 | 12:38:52.8 | -51:09:49 | 0.6523  | 4.8398    | RRab |
| V14 | 592 | 450 | 12:38:36.2 | -51:10:10 | 0.6061  | 4.9464    | RRab |
| V15 | 700 | 334 | 12:38:27.7 | -51:11:26 | 0.6033  | 4.8661    | RRab |
| V16 | 702 | 449 | 12:38:36.2 | -51:11:27 | 0.6270  | 5.1169    | RRab |
| V17 | 729 | 174 | 12:38:15.8 | -51:11:47 | 0.4135  | 4.9608    | EW   |
| V18 | 731 | 565 | 12:38:44.8 | -51:11:47 | 0.6375  | 4.5846    | RRab |

Table 3: Light-curve parameters for the Ru 106 RR Lyr variables. The last column gives the rise-time from minimum to maximum light.

| ID | P[d] | $<V>$ | $A_V$ | $<B>$ | $A_B$ | $<B-V>$ | $(B-V)_{min}$ | $\Delta\phi$ |
|---|---|---|---|---|---|---|---|---|
| V1 | 0.6104 | 17.77 | 0.59 | 18.26 | 0.80 | 0.52 | 0.591 | 0.20 |
| V2 | 0.6268 | 19.08 | 0.51 | 19.51 | 0.72 | 0.51 | 0.589 | 0.20 |
| V5 | 0.6006 | 17.78 | 0.58 | 18.16 | 0.72 | 0.52 | 0.615 | 0.20 |
| V6 | 0.6408 | 17.82 | 0.37 | 18.38 | 0.48 | 0.55 | 0.584 | 0.21 |
| V8 | 0.6208 | 17.75 | 0.49 | 18.23 | 0.71 | 0.50 | 0.564 | 0.17 |
| V9 | 0.5740 | 17.75 | 0.84 | 18.09 | 1.06 | 0.46 | 0.59: | 0.15 |
| V10 | 0.6019 | 17.79 | 0.67 | - | - | 0.56 | 0.595 | 0.16 |
| V12 | 0.6221 | 17.84 | 0.51 | 18.25 | 0.67 | 0.51 | 0.557 | 0.20 |
| V13 | 0.6523 | 17.78 | 0.37 | 18.28 | 0.47 | 0.53 | 0.566 | 0.30 |
| V14 | 0.6061 | 17.88 | 0.77 | 18.28 | 0.94 | 0.49 | 0.557 | 0.17 |
| V15 | 0.6033 | 17.84 | 0.64 | - | - | 0.54 | 0.584 | 0.19 |
| V16 | 0.6270 | 17.80 | 0.47 | 18.32 | 0.62 | 0.53 | 0.581 | 0.19 |
| V18 | 0.6375 | 17.80 | 0.39 | 18.36 | 0.54 | 0.55 | 0.595 | 0.20 |

Table 4: Light-curve parameters for three SX Phe stars and two contact binaries discovered in the field of Ru 106.

| ID | P[d] | $V_{max}$ | $<V>$ | $A_V$ | $(B-V)_{max}$ | $<B-V>$ |
|-----|---------|-------|-------|------|------|------|
| V3 | 0.03952 | 19.88 | 20.05 | 0.30 | 0.22 | 0.36 |
| V4 | 0.04900 | 19.62 | 19.71 | 0.14 | 0.30 | 0.36 |
| V11 | 0.04777 | 20.12 | 20.25 | 0.18 | 0.26 | 0.35 |
| V17 | 0.4135 | 16.60 | 16.78 | 0.48 | 0.84 | 0.88 |
| V7 | 0.2755 | 19.42 | 19.55 | 0.28 | 0.98 | 1.06 |

Table 5: Photometry of variables discovered in the field of Ru 106. The first column lists the heliocentric Julian date of mid-exposure.

| HJD 2449100+ | V V1 | $\sigma$ | V V2 | $\sigma$ | V V3 | $\sigma$ | V V4 | $\sigma$ | V V5 | $\sigma$ | V V6 | $\sigma$ |
|------|------|------|------|------|------|------|------|------|------|------|------|------|
| 4.5463 | 17.620 | 0.012 | 19.041 | 0.023 | 20.064 | 0.035 | 19.745 | 0.032 | 17.622 | 0.033 | 17.593 | 0.009 |
| 4.5504 | 17.606 | 0.014 | 19.066 | 0.021 | 19.850 | 0.028 | 19.729 | 0.044 | 17.653 | 0.075 | 17.597 | 0.017 |
| 4.5612 | 17.556 | 0.014 | 19.103 | 0.019 | 20.057 | 0.029 | 19.730 | 0.036 | 17.660 | 0.046 | 17.627 | 0.023 |
| 4.5655 | 17.525 | 0.009 | 19.088 | 0.016 | 20.140 | 0.037 | 19.713 | 0.026 | 17.561 | 0.016 | 17.608 | 0.014 |

Table 14: Magnitudes, colors, and positions of stars from the field of Ru 106. Rectangular coordinates can be used to identify all stars on the image included in the supplementary data published on the CD-ROM supplemented to the *Astronomical Journal*.

| ID | X | Y | V | $\sigma$ | B-V | $\sigma$ |
|-----|--------|--------|--------|-------|-------|-------|
| 1878 | 130.15 | 213.35 | 20.139 | 0.042 | 1.229 | 0.091 |
| 2652 | 130.54 | 688.01 | 20.691 | 0.061 | 1.039 | 0.111 |
| 2770 | 131.41 | 560.73 | 20.717 | 0.072 | 0.552 | 0.107 |
| 3668 | 131.97 | 691.53 | 21.190 | 0.085 | 1.572 | 0.206 |
| 1741 | 132.38 | 539.53 | 19.987 | 0.035 | 0.788 | 0.059 |

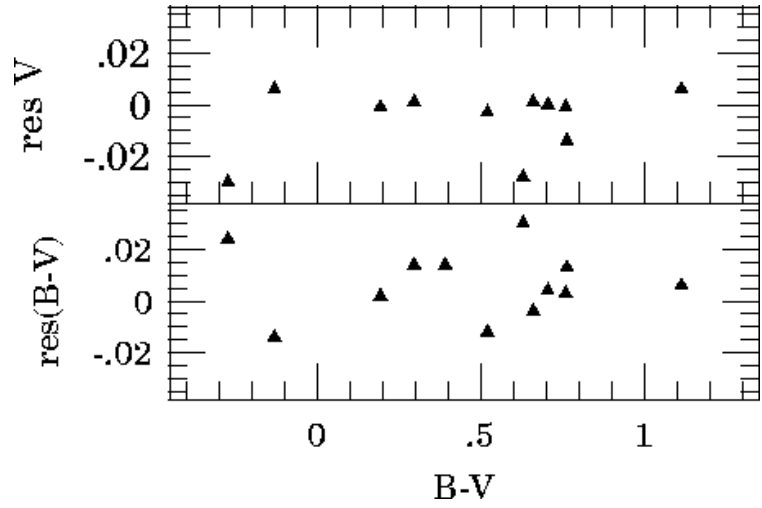

Fig. 1.— Residuals for the Landolt standards observed at the beginning of the night of May 1, 1993.

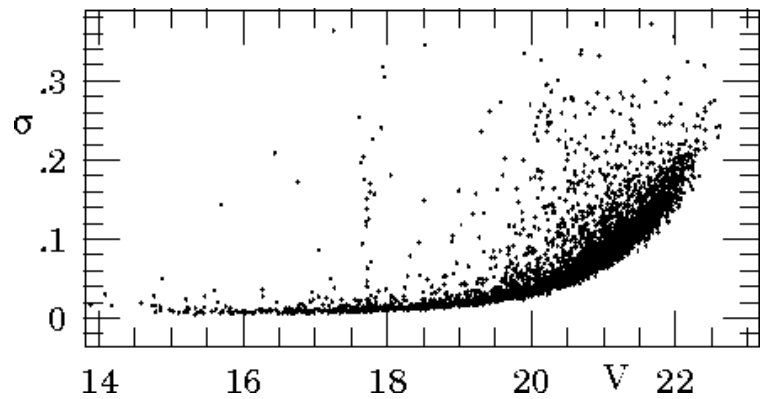

Fig. 2.— Standard deviation versus average V magnitude for stars measured on at least 37 out of 77 frames obtained on the nights of April 27 and 28, 1993. The group of stars with large values of $rms$ present at V $\approx$ 17.8 are RR Lyr variables.

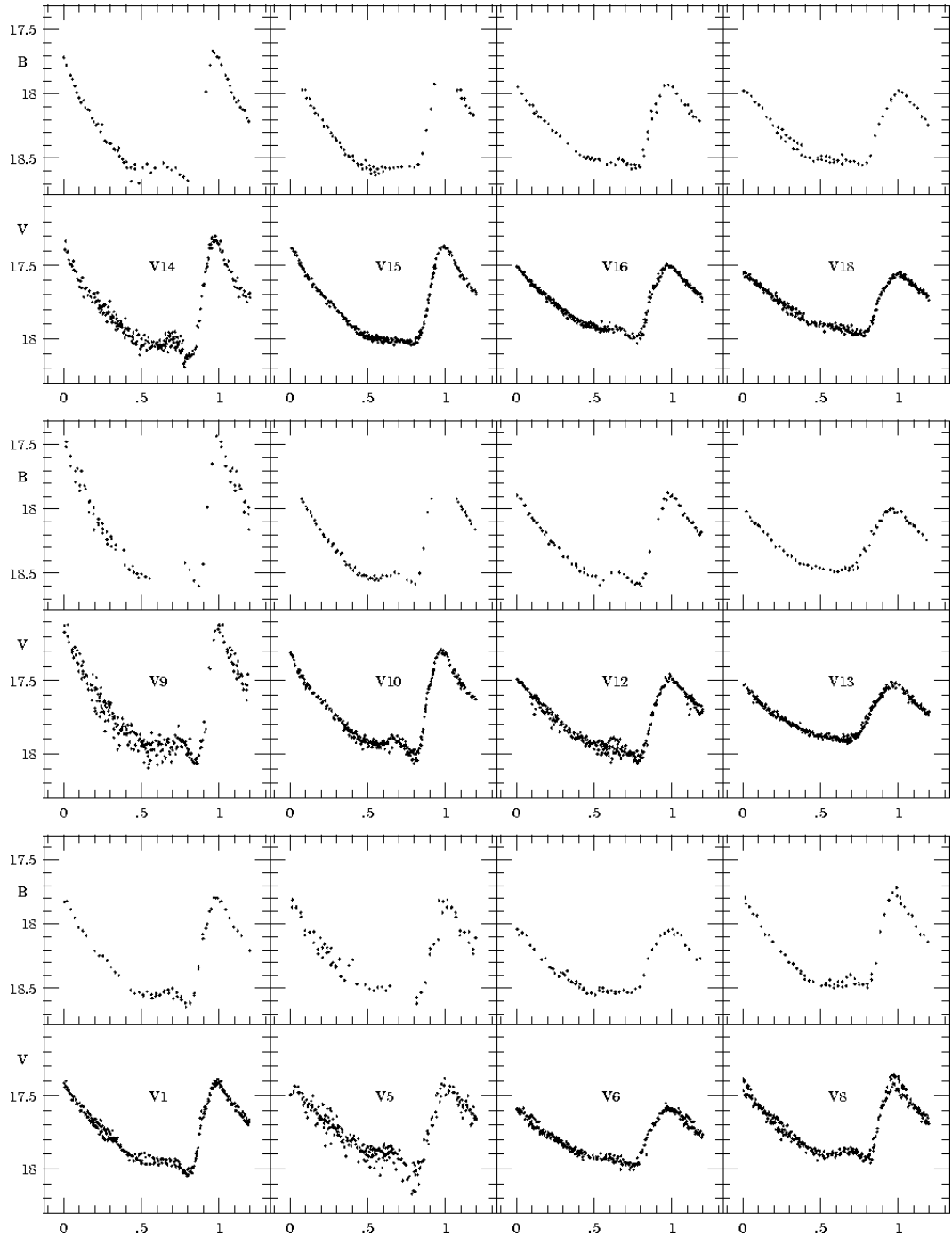

Fig. 3.— The light curves for 12 out of 13 RR Lyr variables discovered in the field of Ru 106. For each star the B-band (upper) and V-band (lower) light curves are shown. Note the same scale of all plots. The light curves for the relatively faint RR Lyr star V2 are shown in Fig. 5.

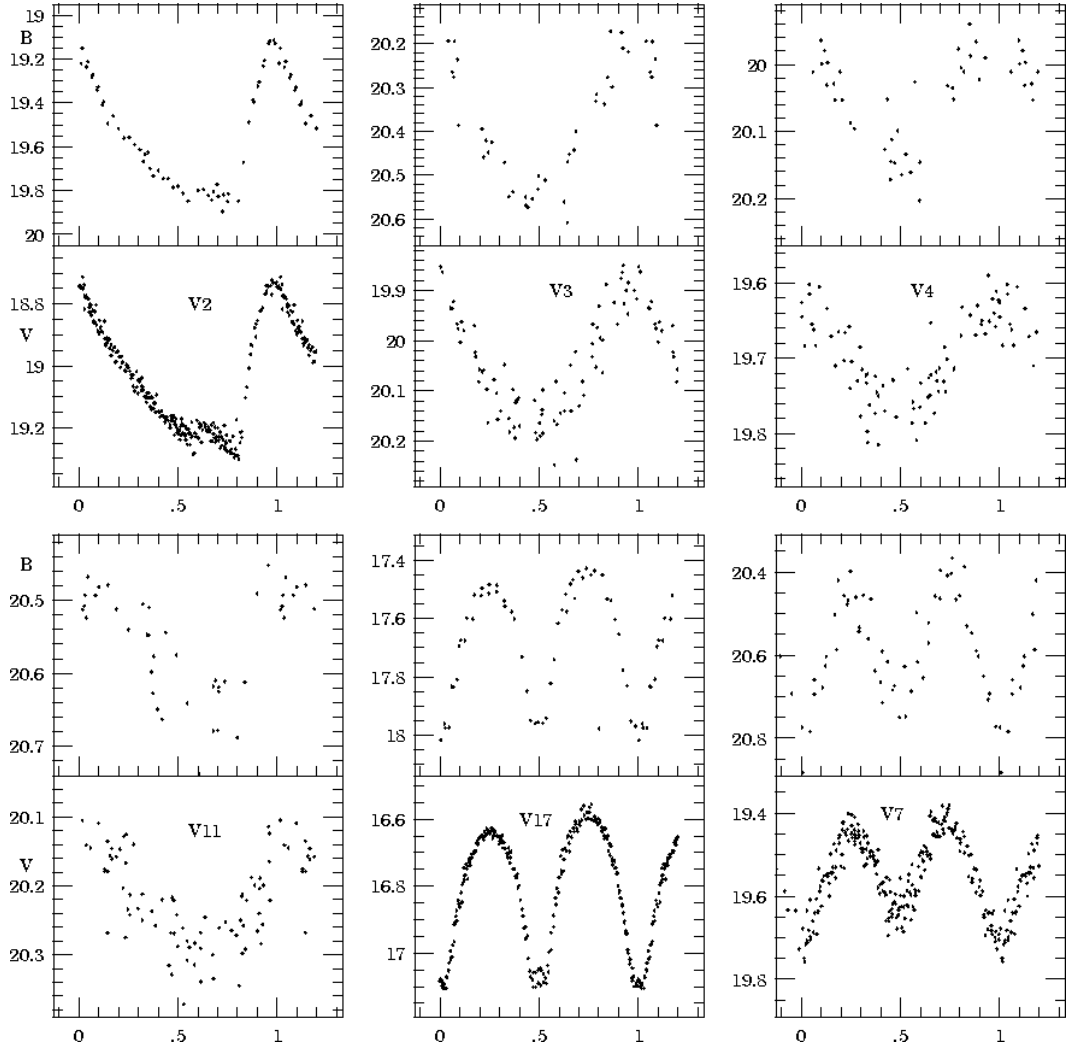

Fig. 4.— The light curves for RR Lyr star V2, SX Phe stars V3, V4 and V11, and contact binaries V17 and V7. For each star the B-band (upper) and V-band (lower) light curves are shown.

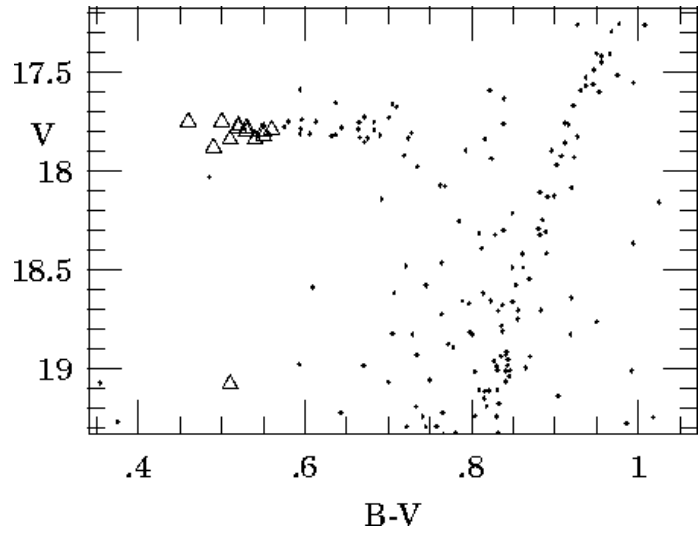

Fig. 5.— A CMD for stars located at distances $R < 2$ arcmin from the center of Ru 106. Positions of RR Lyr variables which were discovered in the whole monitored field are marked with triangles.

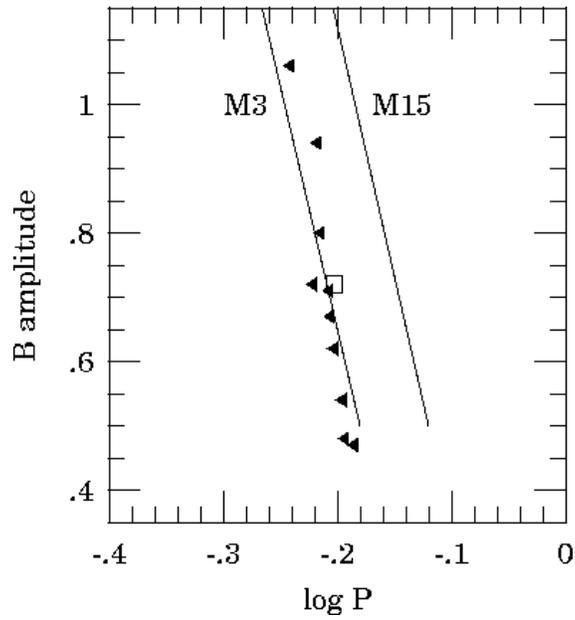

Fig. 6.— The period-amplitude diagram for the Ru 106 RR Lyr variables. The position of a probable field variable V2 is marked with a square. The fiducial lines drawn are for M3 and M15.

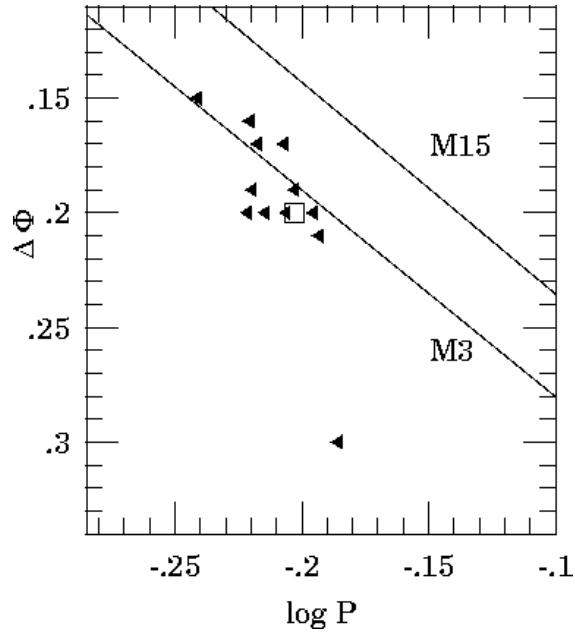

Fig. 7.— Period versus rise-time diagram for RR Lyr variables. The position of a probable field variable V2 is marked with a square. The fiducial lines drawn are for M3 and M15.

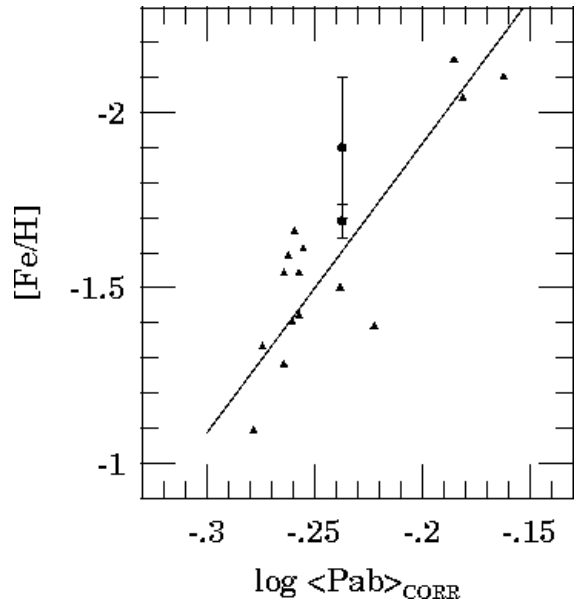

Fig. 8.— Metallicity versus "corrected average period" for 15 clusters rich in RRab variables (Sandage 1993). The relation adopted by Sandage is shown with a continuous line. Two positions of Ru 106 corresponding to [Fe/H] = −1.9 ± 0.2 (Buonanno et al. 1993) and [Fe/H] = −1.69 ± 0.05 (Da Costa et al. 1992) are also marked.

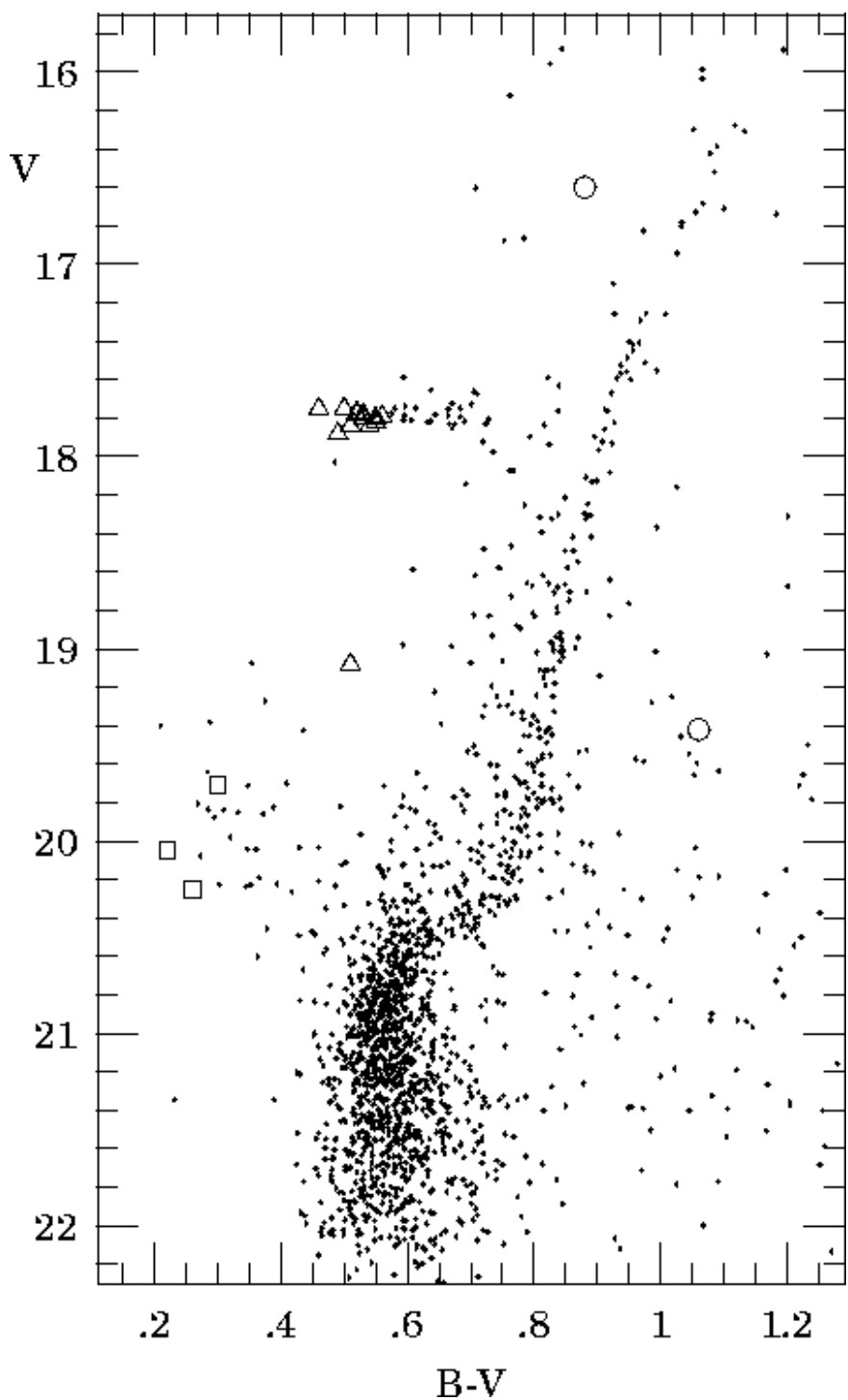

Fig. 9.— The CMD for stars located at distances $R < 2$ arcmin from the center of Ru 106. Positions of variables discovered in the whole monitored field ($11.6 \times 11.6$ arcmin$^2$) are marked with open triangles (RR Lyr), open squares (SX Phe) and open circles (contact binaries).

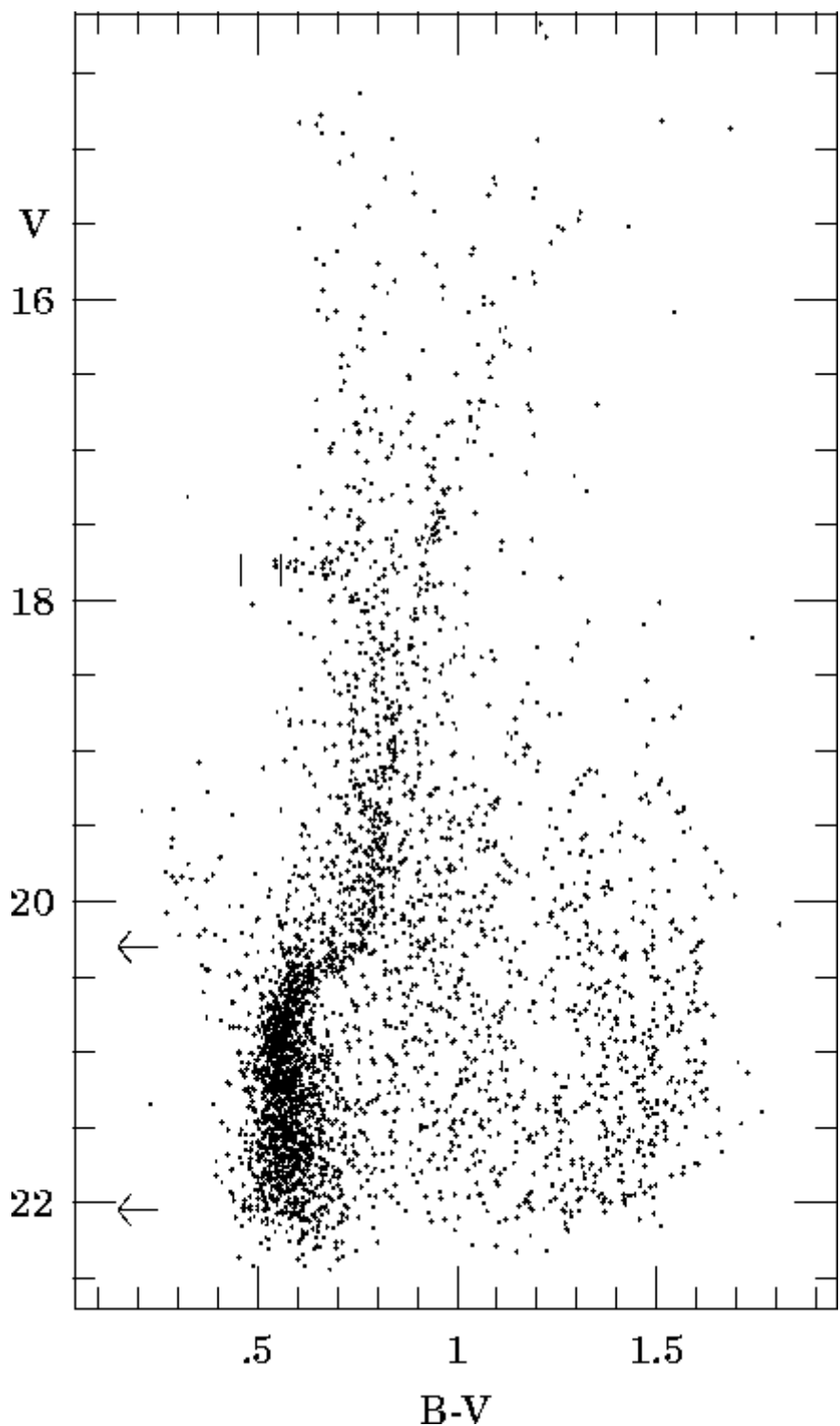

Fig. 10.— The CMD for stars from the field $8.6 \times 8.6$ arcmin$^2$ centered on the center of Ru 106. Identified variables are not marked in this figure. Two vertical bars at $V \approx 17.8$ mark a position of Ru 106 RR Lyr variables. Two horizontal arrows at $V \approx 20.3$ and $V \approx 22.05$ point toward the location of two blue stars with $B - V \approx -0.1$.

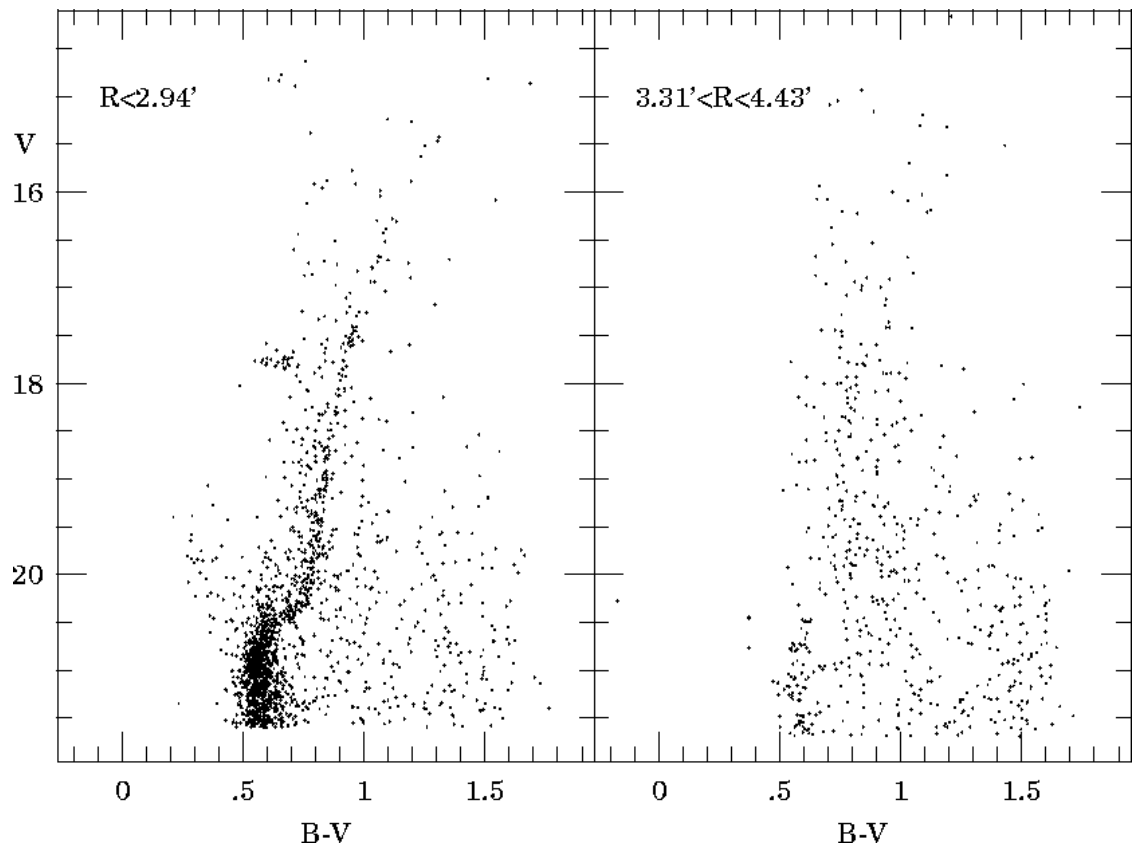

Fig. 11.— The CMDs for the central part of Ru 106 ($R < 2.94$ arcmin) and for the outer ring ($3.31 < R < 4.43$ arcmin). Known variables were not plotted in this figure.

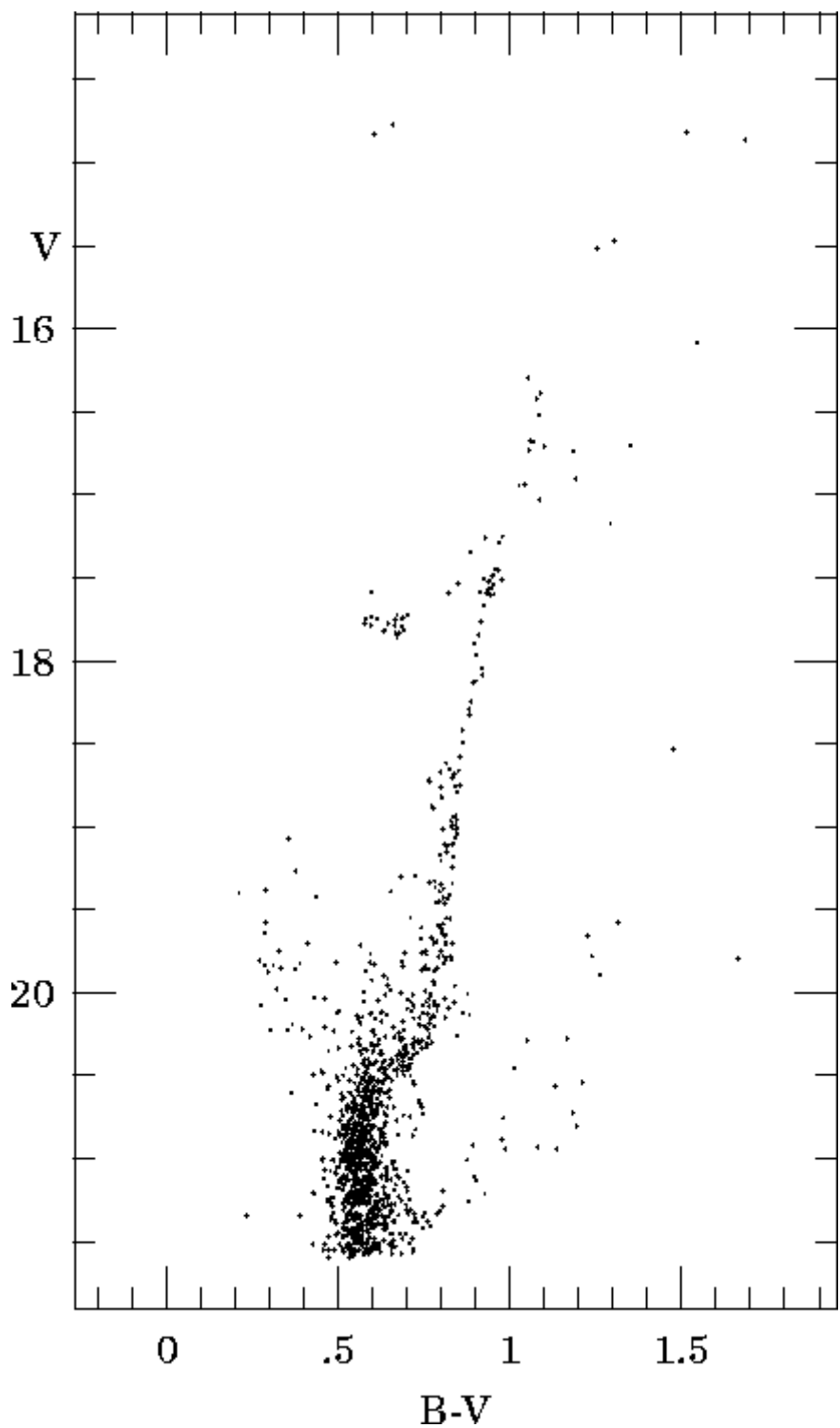

Fig. 12.— The CMD for the central part of Ru 106 ($R < 2.94$ arcmin) where the field stars have been statistically subtracted. Known variables were not plotted in this figure.

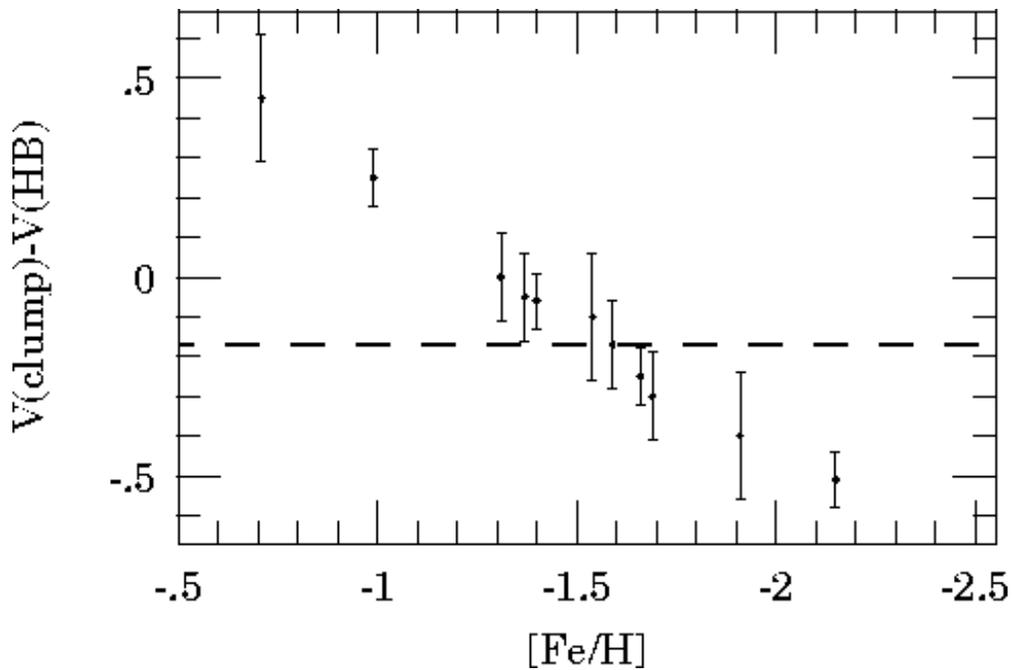

Fig. 13.— The observed values of $\Delta V_{\mathrm{bump}}^{\mathrm{HB}}$ (after Fusi Pecci et al. 1990) as a function of metallicity. The point at [Fe/H] = −2.15 marks the average position for three metal-poor clusters. The dashed line corresponds to $\Delta V_{\mathrm{bump}}^{\mathrm{HB}} = -0.17$ obtained for Ru 106.